\documentclass[aps,prl,superscriptaddress,twocolumn,showpacs]{revtex4-1}

\usepackage{graphicx}
\usepackage{dcolumn}
\usepackage{bm}
\usepackage{amsmath}

\begin{document}

\title{Quantum oscillation of thermal conductivity and violation of Weidemann-Franz law in TaAs$_2$ and NbAs$_2$}

\author{X. Rao}
\affiliation{Department of Physics, Hefei National Laboratory for Physical Sciences at Microscale, and Key Laboratory of Strongly-Coupled Quantum Matter Physics (CAS), University of Science and Technology of China, Hefei, Anhui 230026, People's Republic of China}

\author{X. Zhao}
\email{xiazhao@ustc.edu.cn}

\affiliation{School of Physical Sciences, University of Science and Technology of China, Hefei, Anhui 230026, People's Republic of China}

\author{Y.-Y. Wang}
\affiliation{Department of Physics, Renmin University of China, Beijing 100872, People's Republic of China}
\affiliation{Beijing Key Laboratory of Opto-electric Functional Materials and Micro-nano Devices, Renmin University of China, Beijing 100872, People's Republic of China}

\author{H. L. Che}
\affiliation{Department of Physics, Hefei National Laboratory for Physical Sciences at Microscale, and Key Laboratory of Strongly-Coupled Quantum Matter Physics (CAS), University of Science and Technology of China, Hefei, Anhui 230026, People's Republic of China}

\author{L. G. Chu}
\affiliation{Department of Physics, Hefei National Laboratory for Physical Sciences at Microscale, and Key Laboratory of Strongly-Coupled Quantum Matter Physics (CAS), University of Science and Technology of China, Hefei, Anhui 230026, People's Republic of China}

\author{G. Hussain}
\affiliation{Department of Physics, Hefei National Laboratory for Physical Sciences at Microscale, and Key Laboratory of Strongly-Coupled Quantum Matter Physics (CAS), University of Science and Technology of China, Hefei, Anhui 230026, People's Republic of China}

\author{T.-L. Xia}
\email{tlxia@ruc.edu.cn}
\affiliation{Department of Physics, Renmin University of China, Beijing 100872, People's Republic of China}
\affiliation{Beijing Key Laboratory of Opto-electric Functional Materials and Micro-nano Devices, Renmin University of China, Beijing 100872, People's Republic of China}

\author{X. F. Sun}
\email{xfsun@ustc.edu.cn}

\affiliation{Department of Physics, Hefei National Laboratory for Physical Sciences at Microscale, and Key Laboratory of Strongly-Coupled Quantum Matter Physics (CAS), University of Science and Technology of China, Hefei, Anhui 230026, People's Republic of China}
\affiliation{Institute of Physical Science and Information Technology, Anhui University, Hefei, Anhui 230601, People's Republic of China}
\affiliation{Collaborative Innovation Center of Advanced Microstructures, Nanjing, Jiangsu 210093, People's Republic of China}

\date{\today}

\begin{abstract}

We report a study of thermal conductivity and resistivity at ultra-low temperatures and in high magnetic fields for semi-metal materials TaAs$_2$ and NbAs$_2$ by using single crystal samples. The thermal conductivity is strongly suppressed in magnetic fields, having good correspondence with the large positive magnetoresistance, which indicates a dominant electronic contribution to thermal conductivity. In addition, not only the resistivity but also the thermal conductivity display clear quantum oscillations behavior at subKelvin temperatures and in magnetic fields up to 14 T. The most striking phenomenon is that the thermal conductivity show a $T^4$ behavior at very low temperatures, while the resistivity show a $T$-independent behavior. This indicates a strong violation of the Weidemann-Franz law and points to a non-Feimi liquid state of these materials.

\end{abstract}

\maketitle

Topological materials, such as topological insulators and topological semimetals, have attracted a lot of attention for their exotic physical properties. Band inversion and metallic surface state are two features of topological insulators. While in topological semimetals, band crossing at high symmetry point forms four (two) fold degenerate Dirac (Weyl) points. The typical Dirac semimetal Na$_3$Bi/Cd$_3$As$_2$ and Weyl semimetal \emph{TmPn} (where \emph{Tm} = Ta or Nb, and \emph{Pn} = As or P) exhibit many interesting transport phenomena, such as extremely large magnetoresistance (XMR), chiral anomaly, high mobility and nontrivial Berry phase etc. \cite{J. Xiong, T. Liang, L. P. He, H. Li, C. Shekhar, C. Zhang, J. Hu-1, C.-L. Zhang, F. Arnold} The transition metal dipnictides TaAs$_2$ and NbAs$_2$ are weak topological insulators in the absence of field, but type-II Weyl fermions are suggested to exist if a magnetic field is applied \cite{Y.-Y. Wang, D. S. Wu, Z. Yuan, Y. Luo, B. Shen, D. Gresch}. Both of them exhibit XMR at low temperatures and in high magnetic fields due to the compensated carriers and high mobility \cite{Y.-Y. Wang, D. S. Wu, Z. Yuan, Y. Luo, B. Shen}. In addition, negative longitudinal magnetoresistance has also been observed in them \cite{Y. Luo, B. Shen}, which may be related to the chiral anomaly. In the magnetotransport measurements, the nontrivial band structure is usually expressed with $\pi$ Berry phase which can be extracted from quantum oscillation. While the magnetotransport properties of TaAs$_2$ and NbAs$_2$ have been investigated in details, the heat transport of them still remains to be studied. In this regard, the heat transport at ultra-low temperatures, which is very useful for characterizing the nature of electron system, has actually not been reported for all the known semimetal materials.

The well-known Wiedemann-Franz (WF) law is one of the fundamental properties of a Fermi liquid. It relates the thermal conductivity $\kappa$, the electrical conductivity $\sigma$ and the absolute temperature $T$ through a simple formula, $\kappa/\sigma T = L$, where $L$ is called the Lorentz number and is given by the Sommerfeld's value $L_0 = 2.44\times10^{-8}$ W$\Omega$/K$^2$. It relies on the single-particle description of the transport properties and a purely elastic electron scattering. Although the WF law is usually not obeyed at $T \neq 0$ because of the importance of inelastic scattering, it must be obeyed at $T = 0$ in the Fermi liquid where the electrons are elastically scattered by static disorders. The examination of the WF law at $T \to 0$ by using ultra-low-temperature charge and heat transport measurements can therefore provide a direct judgment on the nature of the electron system. In some previous studies on the semi-metal materials, the WF law was not experimentally confirmed but was usually assumed to be valid \cite{U. Stockert}. In this work, we study the charge and heat transport of TaAs$_2$ and NbAs$_2$ at very low temperatures. It was found that the electronic thermal conductivity exhibits a peculiar $T^4$ behavior at very low temperatures down to several tens of milli-Kelvins, while the resistivity is $T$-independent. This indicates a strong violation of the WF law in these semimetal materials. In addition, the thermal conductivity display clear quantum oscillation in high magnetic fields.

The TaAs$_2$ and NbAs$_2$ single crystals were grown by the chemical vapor transport method. By using the X-ray Laue photograph, the crystals were cut precisely along the crystallographic axes with typical dimensions of $1.7 \times 0.5 \times 0.1$ mm$^3$, where the longest dimension is along the $b$ axis. Resistivity was measured by using the standard four-probe method. Thermal conductivity were measured by using a ``one heater, two thermometers'' technique. Note that these two measurements for each material were done on the same piece of single crystal with the same contacts for direct checking the validity of WF law. These measurements were performed in a $^3$He refrigerator at 300 mK $-$ 30 K and in a $^3$He/$^4$He dilution refrigerator at 70 mK $-$ 1 K, equipped with a 14 T superconducting magnet. The electric and heat currents were applied along the $b$ axis while the magnetic fields were applied perpendicular to the $b$ axis.

\begin{figure}
\centering\includegraphics[clip,width=6.0cm]{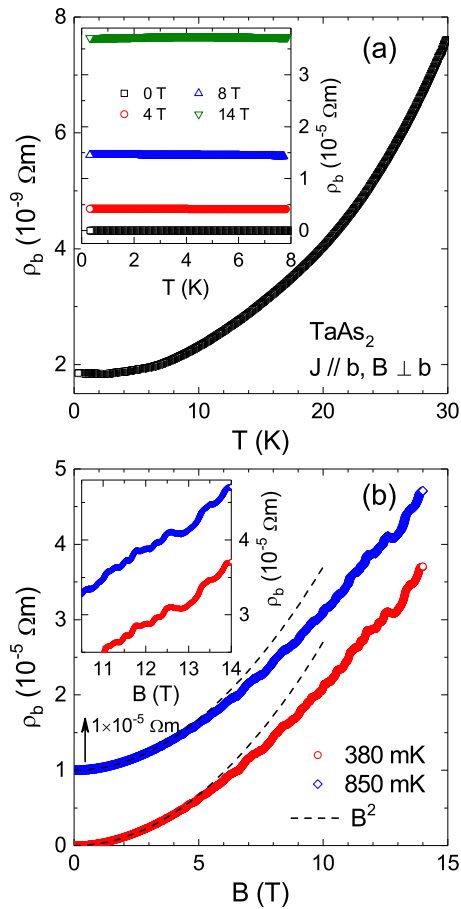}
\caption{Charge transport of a TaAs$_2$ single crystal. The electric current $(J)$ is parallel to the $b$ axis and magnetic field is perpendicular to the $b$ axis. (a) Resistivity as a function of temperature in zero field. The charge transport shows a metallic behavior with a small and $T$-independent residual value at $T \rightarrow 0$ K. Inset: the low-temperature resistivity increases significantly with applying magnetic field (0, 4, 8, and 14 T) and are nearly temperature independent. (b) Magnetic field dependence of resistivity at 380 and 850 mK (the 850 mK data are shifted upward $1\times10^{-5}$ $\Omega$m for clarifying). The dashed lines are quadratic fittings to the low-field data of TaAs$_2$. A semi-classical quadratic behavior at low field and unsaturated positive and extremely large MR at high field indicate the compensation of carriers. Inset: the enlarged part of $\rho(B)$ data at 10$-$14 T, where the SdH oscillations can be seen more clearly.}
\end{figure}

Since TaAs$_2$ and NbAs$_2$ exhibit almost the same experimental results of low-$T$ resistivity and thermal conductivity, we show only the TaAs$_2$ results in the main text. The charge transport of this material had been carefully studied in magnetic fields up to 14 T and at low temperatures down to 2 K \cite{Y.-Y. Wang, Z. Yuan}. Here, the resistivity measurements of a TaAs$_2$ single crystal are carried out at lower temperatures down to 300 mK. In zero magnetic field, the $b$-axis resistivity $\rho_b(T)$ decreases monotonically with lowering temperature until it reaches a finite value (residual resistivity) at $T < 2$ K, as shown in Fig. 1(a), which is characterized as a metallic behavior. With applying magnetic fields ($\perp b$) up to 14 T, the resistivity is strongly enhanced, as shown in the inset of Fig. 1(a). Figure 1(b) shows the field dependence of resistivity at two selected temperatures below 1 K. The $\rho_b(B)$ curves exhibit a semi-classical quadratic behavior at low fields, accompanied with a clear Shubnikov-de Hass (SdH) oscillation at high fields, as shown in the inset of Fig. 1(b). Note that the MR displays unsaturated positive and extremely large value. These phenomena are similar to those observed at temperatures above 2 K and can be understood with the compensated hole and electron of TaAs$_2$ \cite{Y.-Y. Wang, Z. Yuan, M. N. Ali, X. Du, J. M. Ziman}.

\begin{figure}
\centering\includegraphics[clip,width=8.5cm]{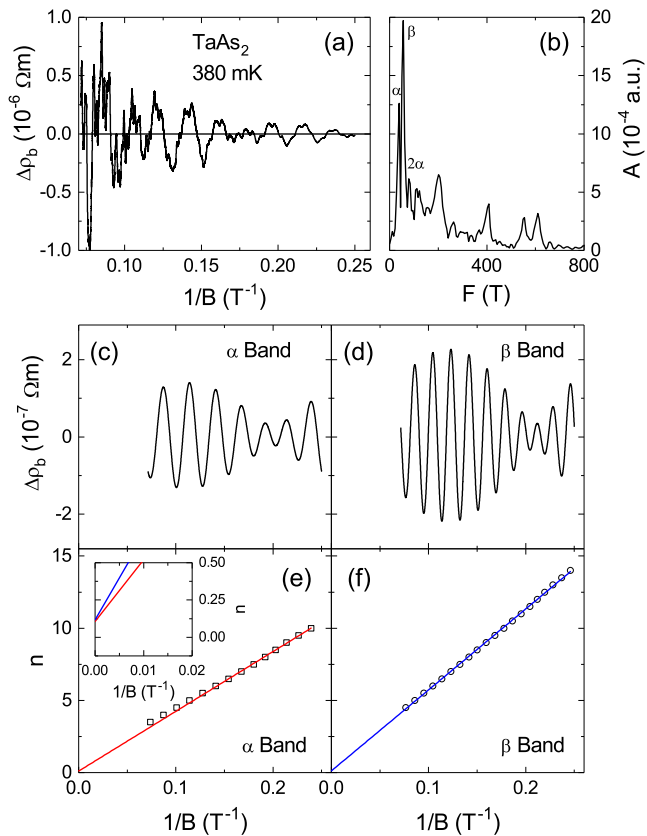}
\caption{Analysis of SdH oscillations at 380 mK in TaAs$_2$. (a) The amplitude of SdH oscillations as a function of the reciprocal of magnetic field. (b) The FFT spectrum of the corresponding SdH oscillation. (c,d) The reconstructed oscillation $\Delta\rho_b$ originating from $F_\alpha$ and $F_\beta$ using inverse FFT. (e,f) Landau index $n$ plotted against $1/B$ for corresponding oscillation frequencies ($F_\alpha$ and $F_\beta$). The lines represent the linear fit to the LL indices obtained at low field, to minimize the influence of the Zeeman effect. Inset: the intercepts of two bands which are in the range 0 to 1/8 (close to 1/8) indicate the nontrivial Berry phase and 3D characteristics for the two different bands.}
\end{figure}

The feature of multiple Fermi surfaces of TaAs$_2$ is verified by the SdH oscillation. After subtracting the high-field background of $\rho_b(B)$ by using a polynomial formula fitting, the SdH oscillation term above 4 T is obtained. Figure 2(a) shows the oscillation amplitude $\Delta\rho_b$ at 380 mK with respect to $1/B$, which forms complicated patterns coming from multiple types of carriers with high mobility \cite{Y.-Y. Wang, Z. Yuan}. This can be seen more clearly by the fast Fourier transform (FFT) analysis. Figure 2(b) presents the FFT analysis on the oscillation data. There are several obvious peaks in the FFT spectra, but the fundamental oscillation frequencies are $F_\alpha = 39$ T and $F_\beta = 56$ T, corresponding to two major Fermi pockets. The complexity of periodic behavior can be attributed to the effect of weak peaks in the FFT spectra. The smaller frequencies obtained in our experiments compared with those from literature \cite{Y.-Y. Wang} are due to the different directions of the magnetic field.

The Berry phase can be obtained from the analysis of SdH oscillations. A nontrivial $\pi$ Berry phase is expected for the relativistic topological fermions \cite{Y. Zhang, H. Murakawa}. The simple way to extract the Berry phase is to map the Landau level (LL) index fan diagrams. The LL index $n$ is linearly dependent on $1/B$, following the Lifshitz-Onsager quantization rule \cite{D. Shoenberg}, and the Berry phase can be extracted from the intercept of the linear extrapolation. As shown in Figs. 2(c, e) and 2(d, f) the LL fan diagrams are used to analyze two major frequencies of SdH oscillation, in which the integer indices and half integer indices are assigned to the peak positions and the valley positions of $\Delta\rho_b$, respectively. To minimize the influence of the Zeeman effect which will split Landau levels, only low field data are used to fit the LL indices \cite{J. Hu-2}. The intercepts are found to be close to zero, which indicates a nontrivial Berry phase for two bands ($\alpha$ and $\beta$) and points to the topological characteristic of Fermi surface.

\begin{figure}
\centering\includegraphics[clip,width=6.0cm]{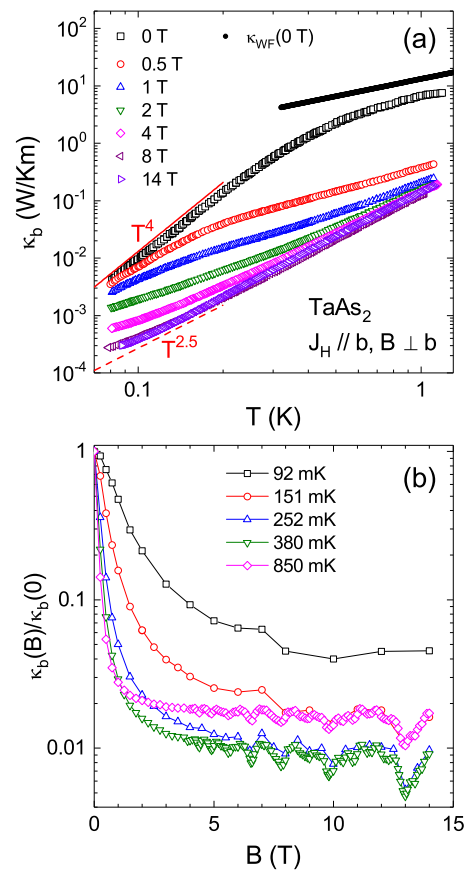}
\caption{Ultra-low-temperature thermal conductivity of the TaAs$_2$ single crystal. The heat current ($J_H$) is parallel to the $b$ axis and magnetic field is perpendicular to the $b$ axis. (a) Thermal conductivity in zero and several different magnetic fields. For comparison, $\kappa_{WF}$(0 T) are calculated data from the zero-field resistivity by using the WF law. (b) Magnetic field dependence of thermal conductivity at some selected temperatures. The 380 and 850 mK curves with dense data points show clear quantum oscillations at high field. Note that the experimental $\kappa_b$ is smaller than $\kappa_{WF}$, in particular at very low temperatures. The solid line indicates a $T^4$ temperature dependence that the $\kappa_b$ follows at the low-temperature limit. The dashed line indicates a $T^{2.5}$ temperature dependence that the $\kappa_b$ roughly follows at the low temperatures and in 14 T field. Since the magnetothermal conductivity behavior indicates the low-temperature thermal conductivity is nearly pure electronic contribution, these phenomena clearly indicate a violation of the WF law.}
\end{figure}

Figure 3 shows the $b$-axis thermal conductivity $\kappa_b$ of TaAs$_2$ measured at very low temperatures down to 70 mK and in magnetic fields ($\perp b$) up to 14 T. The most remarkable phenomenon is that the $\kappa_b$ in zero field follows a $T^4$ behavior at very low temperatures ($< 200$ mK), as shown in Fig. 3(a). This is distinctly different from the usual $T^3$ and $T$ dependence of $\kappa$ for phonons and electrons, respectively, at ultra-low temperatures. With applying magnetic fields, the magnitude of $\kappa_b$ is strongly suppressed, which has good correspondence to the large positive MR effect. This clearly indicates that the main heat carriers are electrons (or holes). At the same time, the field dependence of $\kappa_b$ becomes weaker with increasing field. In the highest field of 14 T, the $\kappa_b$ follows a roughly $T^{2.5}$ behavior at very low temperatures ($< 200$ mK) as shown in Fig. 3(a), which means that the heat transport of electrons (or holes) is significantly suppressed and phonon heat transport remains.

Figure 3(b) shows the magnetic-field dependence of $\kappa_b$ at some selected temperatures. Two notable phenomena are observed. First, the $\kappa_b$ is strongly suppressed by nearly 2 orders of magnitude, as the $\kappa_b(T)$ data also indicate. Apparently, the large negative magneto-thermal conductivity has a direct correlation to the positive large MR and the phonon thermal conductivity should be independent of magnetic field. From this result, it is further confirmed that in zero field the phonon contribution to the thermal conductivity is likely less than several percent. Therefore, the $T^4$ temperature dependence of $\kappa_b$ in zero field is a purely electron transport behavior, which is extremely strange. Second, above 4 T there is clear oscillation behavior sitting on an almost constant background, which seems to be related to the SdH oscillation of the MR data. Note that the quantum oscillation in heat transport has been observed in some other semimetals with thermal conductivity and thermoelectric measurements \cite{M. Matusiak, J. Xiang}, and was usually discussed to be a result of phononic thermal conductivity under the electron-phonon coupling. However, such speculation is lacking of solid physical ground since the MR oscillation is determined by the Fermi energy crossing different Landau levels with the change of applied field. Therefore, the quantum oscillations in thermal conductivity are of purely electronic origin and the high-field constant background is a field-independent phonon transport. In the present work, by carrying out the thermal conductivity measurements with unprecedented low-temperature range for semimetal materials, we reveal the clearest quantum oscillation of electronic thermal conductivity for the first time.

\begin{figure}
\centering\includegraphics[clip,width=8.5cm]{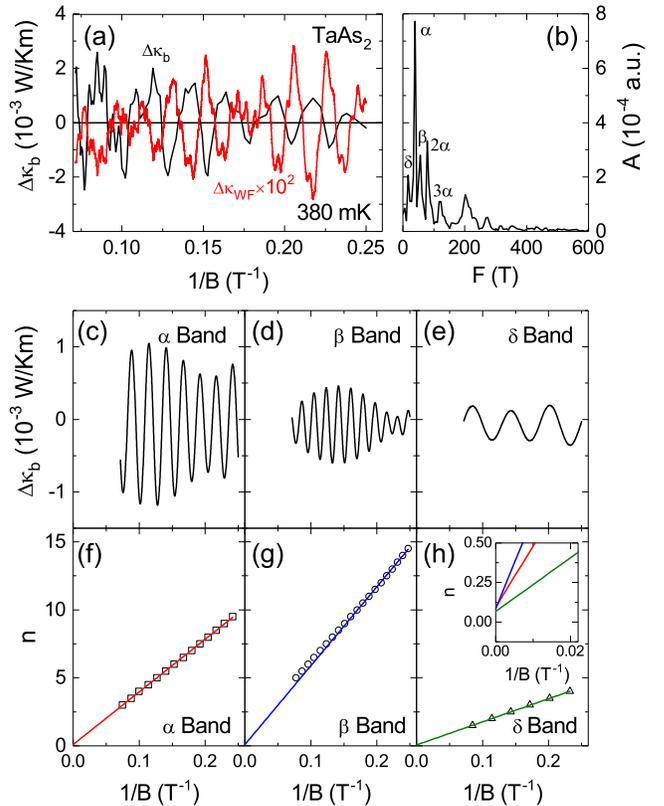}
\caption{Analysis of quantum oscillations of thermal conductivity at 380 mK in TaAs$_2$. (a) The amplitude of quantum oscillations of thermal conductivity as a function of the reciprocal of magnetic field. For comparison, $\Delta\kappa_{WF}$ represents the calculated oscillation term from the resistivity SdH oscillation by using the WF law. Note that the phase of quantum oscillations of experimental $\kappa_b(B)$ is shifted by $\pi/2$ in relation to the SdH oscillation. In addition, the $\Delta\kappa_b(B)$ oscillation is about 100 times larger than the calculated $\Delta\kappa_{WF}$. (b) The FFT spectrum of quantum oscillations of thermal conductivity. (c,d,e) The reconstructed oscillations $\Delta\kappa_b$ originated from $F_\alpha$, $F_\beta$ and $F_\delta$ using inverse FFT. (f,g,h) Landau index $n$ plotted against $1/B$ for corresponding oscillation frequencies ($F_\alpha$, $F_\beta$ and $F_\delta$). The lines represent the linear fits to the LL indices obtained at low field to minimize the influence of the Zeeman effect. Inset: the intercepts of three bands which are in the range 0 to 1/8 (close to 1/8) indicate the nontrivial Berry phase and 3D characteristic for three different bands, similar to the results of the resistivity.}
\end{figure}

The quantum oscillation of thermal conductivity can be analyzed in the similar way to that of MR. After subtracting an appropriate background with a polynomial fitting from the corresponding data, the oscillatory part of thermal conductivity is obtained. As a representative result, Fig. 4(a) presents the oscillation of thermal conductivity at 380 mK. The FFT spectrum of the quantum oscillation in the thermal conductivity data is shown in Fig. 4(b). It is found that there are three frequencies related to the Fermi pockets, labeled as $F_\alpha$, $F_\beta$, and $F_\delta$. Among them, $F_\alpha = 39$ T and $F_\beta = 56$ T are well resolved and are comparable with those obtained by resistivity measurements, while the smaller frequency $F_\delta$ (= 16.8 T) can not be detected unambiguously from the resistivity measurements due to the different experimental resolution of each technique. According to the Onsager relation, the frequency $F$ is proportional to the cross sectional area $A_F$ of the Fermi surface normal to the magnetic field \cite{D. Shoenberg}. Therefore, the small oscillation frequency $F_\delta$ may originate from the Fermi pocket much smaller than those related to $F_\alpha$ and $F_\beta$.

The nature of topological fermions participating in quantum oscillations of thermal conductivity can be revealed from further quantitative analysis. Before analyzing the LL index fan diagrams through the oscillation of thermal conductivity, it should be noted that the phase of quantum oscillations of $\Delta\kappa_b(B)$ is shifted by $\pi/2$ in relation to the SdH oscillations. To display clearly this phase shift, the experimental oscillation of $\Delta\kappa_b(B)$ is compared with the calculated $\Delta\kappa_{WF}(B)$ from the SdH oscillation by using the WF law, as shown in Fig. 4(a). This phase shift and two-order of magnitude difference between $\Delta\kappa_b(B)$ and $\Delta\kappa_{WF}(B)$ actually again indicate that the WF law is invalid for this material although it exhibits good conducting properties at low temperatures. Therefore, the integer LL indices and half integer indices should be assigned to the valley positions and the peak positions of $\Delta\kappa_b$, which is opposite to the case of $\Delta\rho_b$.

As shown in Figs. 4(c--h), we use LL fan diagrams to analyze three major oscillation frequencies of thermal conductivity. To minimize the influence of the Zeeman effect, only the low field data are used to fit LL indices \cite{J. Hu-2}. All the intercepts are close to zero, indicating a nontrivial Berry phase for the three different bands ($\alpha$, $\beta$ and $\delta$). The nontrivial Berry phases from thermal conductivity also reflect the topological characteristic of Fermi surfaces.

To summarize the main experimental results of TaAs$_2$: (i) In zero field, the low-temperature thermal conductivity is dominated by charge carriers, whose transport ability is significantly suppressed in high magnetic fields. (ii) The low-temperature thermal conductivity displays a quantum oscillation behavior in high magnetic fields that has some correspondence to the SdH oscillations. (iii) The most important finding is a $T^4$-dpendence of electron thermal conductivity in zero field and at very low temperatures. Since the resistivity is a simple $T$-independent behavior at very low temperatures (residual resistivity), the WF law is clearly violated in this semimetal material.

The WF law is commonly not obeyed at finite temperatures due to the inelastic scattering \cite{J. M. Ziman}, as observed in elemental metals (electron-phonon scattering), Weyl semimetals and in correlated metals (electron-electron scattering) \cite{G. K. White, J. Paglione, A. Jaoui}. In contrast, the strong violation of the WF law at zero-temperature limit in a metallic system points to a breakdown of the Fermi liquid. In some heavy-fermion material, a breakdown of the WF law at zero-temperature limit was found near a magnetic quantum critical point \cite{H. Pfau}, although the later study indicated very high requirements for the experimental testing of the WF law in such case \cite{M. Taupin}. It is notable that the violation of the WF law has been found in some particular case for Fermi-liquid system. In graphene, the breakdown of the WF law was reported and was attributed to hydrodynamic effect \cite{J. Crossno, A. Lucas}. This could arise from the electron-electron interaction within Fermi-liquid paradigm and might not imply any non-Fermi-liquid behavior \cite{A. Lucas}. However, the present work displays a strong violation of the WF law that cannot be explained by the hydrodynamic effect in Fermi liquid, considering the peculiar $T^4$ behavior of electronic thermal conductivity. The theoretical description for the violation of the WF law and the nature of these semimetals is called for.

\begin{acknowledgements}

We thank H. M. Weng and K. Yang for helpful discussions. This work was supported the National Natural Science Foundation of China (Grants No. 11574286, No. 11574391, No. U1832209, and No. 11874336), the National Basic Research Program of China (Grants No. 2015CB921201 and No. 2016YFA0300103) and Users with Excellence Project of Hefei Science Center CAS (Grant No. 2018HSC-UE012). T.L.X. is also supported by the Fundamental Research Funds for the Central Universities, and the Research Funds of Renmin University of China (Grants No. 14XNLQ07 and No. 18XNLG14).

\end{acknowledgements}

\end{document}